\documentstyle[amssymb,aps,prd,twocolumn]{revtex}
\begin{document}
\title{Casimir energy in multiply connected static hyperbolic Universes}
\author{Daniel M\"uller\thanks{%
Electronic Address: muller@fis.unb.br}}
\address{{\it Instituto de F\'{i}sica, Universidade de Bras\'{i}lia}\\
{\it Campus Universit\'{a}rio Darcy Ribeiro, Caixa Postal 04455,}\\
{\it CEP }70919-970, {\it Bras\'{i}lia, DF, Brazil}}
\author{Helio V. Fagundes\thanks{%
Electronic Address: helio@ift.unesp.br}}
\address{{\it Instituto de F\'{i}sica Te\'{o}rica, Universidade Estadual Paulista}\\
{\it Rua Pamplona, 145, CEP 01405-900, S\~{a}o Paulo, SP, Brazil}}
\author{Reuven Opher\thanks{%
Electronic Address: opher@astro.iag.usp.br}}
\address{{\it Instituto de Astronomia, Geof\'{i}sica e Ci\^{e}ncias Atmosf\'{e}ricas,}%
\\
Universidade de S\~{a}o Paulo\\
{\it Rua do Mat\~{a}o, 1226, Cidade Universit\'{a}ria, CEP 05508-900, }\\
{\it S\~{a}o Paulo, SP, Brazil}}
\maketitle

\begin{abstract}
We generalize a previously obtained result, for the case of a few other
static hyperbolic universes with manifolds of nontrivial topology as $\,\,$%
spatial sections.

\centerline{PACS numbers: 98:80Cq, 98:80Hw}
\end{abstract}

\input epsfig.sty

\section{Introduction}

As is well known, Einstein equations (EQ) restrict the local geometry of
spatially homogeneous and isotropic spacetimes, to those of $R^{3}$, $S^{3}$%
, or $H^{3}$. Recent observational data indicate that the curvature of the
universe is small, without ruling out the case of negative curvature. On the
other hand, the EQ are insensitive to a global nontrivial topology of space,
which can be a compact hyperbolic 3-space ${\cal M}$, which is isometric to
a quotient space $H^{3}/\Gamma $. Here $\Gamma $ is a nontrivial discrete
group of isometries ($\,$known as holonomy group), which acts freely and
properly discontinuously on the covering space $H^{3}.$ Also important is
the fact that $\Gamma $ is isomorphic to the fundamental group $\pi _{1}(%
{\cal M)}$, which is a group of homotopy classes of maps of the circle $S^{1}
$ into ${\cal M}$ \cite{dfn2}. Since $\pi _{1}({\cal M)}$ is nontrivial, $%
{\cal M}$ is multiply connected.

The 3-space ${\cal M}$ may be represented by a fundamental polyhedron $FP$ \
in $H^{3},$ with an even number of faces, whose copies $\gamma (FP)$, $%
\gamma \in \Gamma ,$ fill up the entire $H^{3}$. The faces of $FP$ are
pairwise identified by the basic elements, or generators, of $\ \Gamma $.
The resulting manifold is a bundle with discrete fibers $\Gamma p$ over base
points $p$ in the fundamental polyhedron. These fibers are the points of the
quotient space. \ \ \ 

Among the first applications of the topology considerations, there was an
attempt to explain multiple quasar images \cite{fagundes}.

For recent reviews of topology in connection with cosmology, see \cite{jl}, 
\cite{LR}, and the article \cite{glluw} for compactifications of the
3-sphere.

The first astrophysical limits on the topology of the universe were obtained
for a 3-torus $T^{3}$. Accordance with the homogeneity of the CMBR puts a
lower limit on the size of the fundamental cell, about $3000$ Mpc, which is
a cube in the cases of \cite{sok} and \cite{as}. Later on, it was shown that
this result is very sensitive to the type of the compactifications of the
spatial sections. For a universe with spatial sections $T^{2}\times R$, the
fundamental cell's size is about $1/10$ of the horizon, and is compatible
with the homogeneity of the cosmic microwave background radiation (CMBR) 
\cite{rouk}.

In compact universes, the pair separation histogram would present spikes for
characteristic distances. At first it was thought that this technique, known
as the {\it crystallographic method}, was able to totally determine the
topology of the universe \cite{lll}. It turned out that the crystallographic
method only applied when the holonomy group contained at least a Clifford
translation, i.e. a translation which moves all the points by the same
distance \cite{llu} and \cite{gtrb}. Generalizations of the crystallographic
method were proposed, for example in \cite{fg}.

Also in compact universes the light front of the CMBR interacts with itself
producing circles in its sky pattern \cite{css2}.

A recent result has called our attention to the possibility that methods
based on multiple images will prove not to be efficient \cite{grt}. The
reasoning is that, according to observations, the curvature is very small,
so the fundamental regions are so big that there has not been time enough
for the formation of ghost images. The result is that for low curvature
universes such as ours, only compact universes with the smallest volumes
could be detected by pattern repetitions.

A very attractive argument in favor of compact hyperbolic manifolds is
related to pre-inflationary homogenization through chaotic mixing \cite{css3}%
. The effect is the same that arises in compact hyperbolic surfaces. The
geodesic motion on a surface of genus $g>1$ shows the absence of KAM torus 
\cite{17}. Not only it is ergodic but also satisfies the Anosov property,
which indicates the presence of strong chaos \cite{17}. The chaotic
properties of the $g=2$ torus were previously studied, for example in \cite
{lb}.

We extend our previously obtained result \cite{dhr} (see also \cite{dh}) for
a few more compact hyperbolic universes. In \cite{dhr}, we numerically
calculated the Casimir effect of a scalar field for a static universe whose
spatial section is the Weeks manifold, the smallest volume (with curvature
normalized to $K=-1$) compact hyperbolic manifold known. The outcome of our
calculation is in fact a Casimir energy density $\rho _{C}$.

In compactifications of flat space, the strictly speaking Casimir energy can
be analytically obtained through the very elegant zeta functions techniques;
for a recent review see \cite{most}. Unfortunately this beautiful formalism
does not yield an analytical result in the case of hyperbolic compact
manifolds, since the spectrum of the Laplace-Beltrami operator can only be
determined numerically.

We use the point splitting technique in the covering space, which is a
static hyperbolic universe. The obtained propagator is exact and possesses
information about the global properties of the manifold, in the sense that
the infrared modes are taken into account.

When the spacetime is multiply connected, the propagator is obtained as the
usual sum over paths: all geodesics connecting the two points are taken into
account.

We find a static hyperbolic solution for the EQ in section II. In section
III, we write the expression for the vacuum expectation value of the
energy-momentum tensor in compact hyperbolic universes. We obtain in section
IV the numerical values of the Casimir energy density in a few multiply
connected static spacetimes. Our conclusions are presented in section V. (We
use natural units, $G=c=\hbar =1$, except in section II.)

\section{Quantum Field Theory in The Static Universe $R\times H^3$ \label{s1}%
}

The hyperbolic space sections $H^{3}$, can be realized as a surface 
\begin{equation}
\left( x-x^{\prime }\right) ^{2}+\left( y-y^{\prime }\right) ^{2}+\left(
z-z^{\prime }\right) ^{2}-\left( w-w^{\prime }\right) ^{2}=-a^{2},
\label{cond}
\end{equation}
imbedded in a Minkowski 4-space 
\[
dl^{2}=dx^{2}+dy^{2}+dz^{2}-dw^{2}.
\]
As this space is homogeneous, we explicitly write the origin of coordinates $%
(x^{\prime },y^{\prime },z^{\prime },w^{\prime })$. It can easily be seen
that its isometry group is the proper, orthocronous Lorentz group $%
SO^{\uparrow }(1,3),$ which is isomorphic to $PSL(2,C)=SL(2,C)/\{\pm I\}$ 
\cite{dfn}. With the constraint of Eq. (\ref{cond}) on the line element we
obtain 
\begin{eqnarray}
dl^{2} &=&dx^{2}+dy^{2}+dz^{2}  \nonumber \\
&&-\frac{\left( \left( x-x^{\prime }\right) dx+\left( y-y^{\prime }\right)
dy+\left( z-z^{\prime }\right) dz\right) ^{2}}{\left( x-x^{\prime }\right)
^{2}+\left( y-y^{\prime }\right) ^{2}+\left( z-z^{\prime }\right) ^{2}+a^{2}}%
,  \nonumber \\
ds^{2} &=&-dt^{2}+dl^{2}=g(x,x^{\prime })_{\mu \nu }dx^{\mu }dx^{\nu },
\label{elxp}
\end{eqnarray}
where we interchangeably write $\left( x^{0},x^{1},x^{2},x^{2}\right)
\longleftrightarrow \left( t,x,y,z\right) $. Both connections, $\nabla _{x}$
and $\nabla _{x^{\prime }}$, compatible with the metric of Eq. (\ref{elxp}),
can be defined through \ 
\begin{eqnarray}
&&\nabla _{\mu }g(x,x^{\prime })_{\alpha \beta }\equiv 0\ ,  \label{cn1} \\
&&\nabla _{\mu ^{\prime }}g(x,x^{\prime })_{\alpha \beta }\equiv 0\ .
\label{cn2}
\end{eqnarray}
The expression in Eq. (\ref{elxp}) is the popular Robertson-Walker line
element, which written in the Lobatchevsky form reads 
\begin{equation}
ds^{2}=-dt^{2}+a^{2}\left[ d\chi ^{2}+\sinh ^{2}\chi \left( d\theta
^{2}+\sin ^{2}\theta d\phi ^{2}\right) \right] \ ,  \label{ell}
\end{equation}
with \ 
\[
\sinh ^{2}\chi =\frac{\left( x-x^{\prime }\right) ^{2}+\left( y-y^{\prime
}\right) ^{2}+\left( z-z^{\prime }\right) ^{2}}{a^{2}}\ .
\]

As is well known, the EQ for the homogeneous and isotropic space sections in
Eq. (\ref{ell}), with $a=a(t)$, reduce to the Friedmann-Lema\^{i}tre
equations 
\begin{eqnarray*}
&&\left( \frac{\dot{a}}{a}\right) ^{2}-\frac{1}{a^{2}}=\frac{8\pi G}{3}\rho +%
\frac{\Lambda }{3}\ , \\
&&2\left( \frac{\ddot{a}}{a}\right) +\left( \frac{\dot{a}}{a}\right) ^{2}-%
\frac{1}{a^{2}}=-8\pi Gp+\Lambda \ ,
\end{eqnarray*}
where the right-hand side comes from the classical energy-momentum source
for the geometry, $T^{\mu \nu }=(\rho +p)u^{\mu }u^{\nu }+pg^{\mu \nu }$,
plus the cosmological constant term $\Lambda g^{\mu \nu }$.

We assume that the universe was radiation dominated near the Planck era,
hence $p=\rho /3$, and obtain the following static solution 
\begin{eqnarray}
&&a=\sqrt{\frac{3}{2|\Lambda |}}\ ,  \nonumber \\
&&\rho =\frac{\Lambda }{8\pi G}\ ,  \nonumber \\
&&ds^{2}=-dt^{2}+a^{2}\left[ d\chi ^{2}+\sinh ^{2}\chi \left( d\theta
^{2}+\sin ^{2}\theta d\phi ^{2}\right) \right] \ ,  \label{elpc}
\end{eqnarray}
where the cosmological constant is negative.

We now wish to evaluate the vacuum expectation value of the energy density
for the case of a universe consisting of a classical radiation fluid, a
cosmological constant, and a non-interacting quantum scalar field $\phi $.
The solution of EQ is given in Eq. (\ref{elpc}), where the quantum back
reaction is disregarded. We use the point splitting method in the universal
covering space $R\times H^{3}$, for which the propagator is exact. The point
splitting method was constructed to obtain the renormalized (finite)
expectation values of the quantum mechanical operators. It is based on the
Schwinger formalism \cite{Schwinger}, and was developed in the context of
curved space by DeWitt \cite{dWitt}. Further details are contained in the
articles by Christensen \cite{chris1}, \cite{chris2}. For a review, see \cite
{GMM}.

Metric variations in the scalar action 
\[
S=-\frac{1}{2}\int \sqrt{-g}(\phi _{,\rho }\phi ^{,\rho }+\xi R\phi
^{2}+m^{2}\phi ^{2})d^{4}x\ ,
\]
with conformal coupling $\xi =1/6,$ give the classical energy-momentum
tensor 
\begin{eqnarray}
T_{\mu \nu } &=&\frac{2}{3}\phi _{,\mu }\phi _{,\nu }-\frac{1}{6}\phi
_{,\rho }\phi ^{,\rho }g_{\mu \nu }-\frac{1}{3}\phi \phi _{;\mu \nu } 
\nonumber \\
&&+\frac{1}{3}g_{\mu \nu }\phi \Box \phi +\frac{1}{6}G_{\mu \nu }\phi ^{2}-%
\frac{1}{2}m^{2}g_{\mu \nu }\phi ^{2}\ ,  \label{tmunu}
\end{eqnarray}
where $G_{\mu \nu }$ is the Einstein tensor. As expected for massless
fields, it can be verified that the trace of the above tensor is identically
zero if $m=0.$ Variations with respect to $\phi $ result in the curved space
generalization of the Klein-Gordon equation, 
\begin{equation}
\square \phi -\frac{R}{6}\phi -m^{2}\phi =0\ .  \label{ekg}
\end{equation}

The renormalized energy-momentum tensor involves field products at the same
spacetime point. Thus the idea is to calculate the averaged products at
separate points, $x$ and $x^{\prime }$, taking the limit $x^{\prime
}\rightarrow x$ in the end. 
\begin{equation}
\langle 0|T_{\mu \nu }\left( x\right) |0\rangle =\lim_{x^{\prime
}\rightarrow x}T(x,x^{\prime })_{\mu \nu }\ ,  \label{Tnl}
\end{equation}
with

\begin{eqnarray}
&&T(x,x^{\prime })_{\mu \nu }  \nonumber \\
&=&\left[ \frac{1}{6}\left( \nabla _{\mu }\nabla _{\nu ^{\prime }}+\nabla
_{\mu ^{\prime }}\nabla _{\nu }\right) -\frac{1}{12}g(x)_{\mu \nu }\nabla
_{\rho }\nabla ^{\rho ^{\prime }}\right.  \nonumber \\
&&\left. \left. -\frac{1}{12}\left( \nabla _{\mu }\nabla _{\nu }+\nabla
_{\mu ^{\prime }}\nabla _{\nu ^{\prime }}\right) +\frac{1}{48}g(x)_{\mu \nu
}\left( \square +\square ^{\prime }\right) \right. \right.  \label{Tmn} \\
&&\left. \left. +\frac{1}{12}\left( R(x)_{\mu \nu }-\frac{1}{4}R(x)g(x)_{\mu
\nu }\right) -\frac{1}{8}m^{2}g(x)_{\mu \nu }\right] \right.  \nonumber \\
&&G^{(1)}(x,x^{\prime })\ ,  \nonumber
\end{eqnarray}
where the covariant derivatives are defined in Eqs. (\ref{cn1}) and (\ref
{cn2}), and $G^{(1)}$ is the Hadamard function, which is the expectation
value of the anti-commutator of $\phi (x)$ and $\phi (x^{\prime })$ (see
below). We stress that the quantity $T(x,x^{\prime })_{\mu \nu }\ $only
makes sense after the limit in Eq. \ref{Tnl} is taken. The geometric
quantities such as the metric and the curvature are regarded as classical
entities. $g(x)_{\mu \nu }=g(x,x^{\prime }=0)_{\mu \nu }$ is obtained from
the line element in Eq. (\ref{elxp}).

The causal Green function, of Feynman propagator, is obtained as 
\[
G(x,x^{\prime })=i\langle 0|T\phi (x)\phi (x^{\prime })|0\rangle \ ,
\]
where $T$ is the time-ordering operator. Taking its real and imaginary
parts, 
\begin{equation}
G(x,x^{\prime })=G_{s}(x,x^{\prime })+\frac{i}{2}G^{(1)}(x,x^{\prime })\ ,
\label{dfg}
\end{equation}
we get for the Hadamard function 
\[
G^{(1)}(x,x^{\prime })=\langle 0|\{\phi (x),\phi (x^{\prime })\}|0\rangle \
=2%
%TCIMACRO{\func{Im}}%
%BeginExpansion
\mathop{\rm Im}%
%EndExpansion
G(x,x^{\prime })\ .
\]

\section{The Feynman Propagator and the Casimir Energy Density in $R\times 
{\cal M}$}

Green functions, as any other function defined in the spatially compact
spacetime $R\times {\cal M}$, must have the same periodicities of the
manifold ${\cal M}$ itself. One way of imposing this periodicity is by
determining the spectrum of the Laplacian, which can only be done
numerically.

Another method imposes the periodicity by brute force, 
\[
f_{{\cal M}}(x)=\sum_{\gamma \in \Gamma }f(\gamma x)\ .
\]
The above expression is named the Poincar\'{e} series, and when it
converges, it defines functions $f_{{\cal M}}$ on the manifold ${\cal M}$.

We define the operator 
\begin{equation}
F(x,x^{\prime })=F(x)/\sqrt{-g}\delta (x-x^{\prime })\ ,  \label{okf}
\end{equation}
where $F(x)=\square -R/6-m^{2},$ and introduce an auxiliary evolution
parameter $s$ and a complete orthonormal set of states $|x\rangle $, such
that 
\begin{eqnarray}
&&G(x,x^{\prime })=\langle x|\hat{G}|x^{\prime }\rangle \ ,  \nonumber \\
&&F(x,x^{\prime })=\langle x|\hat{F}|x^{\prime }\rangle \ ,  \nonumber \\
&&\hat{G}=i\int_{0}^{\infty }e^{-is\hat{F}}ds\ .  \label{iokf}
\end{eqnarray}
This last equation implies that $\hat{G}=(\hat{F}-i0)^{-1}$, hence the
causal Green function becomes 
\begin{equation}
G(x,x^{\prime })=i\int_{0}^{\infty }ds\langle x|\exp (-is\hat{F})|x^{\prime
}\rangle \ ,  \label{int}
\end{equation}
and the matrix element $\langle x|\exp (-is\hat{F})|x^{\prime }\rangle
=\langle x(s)|x^{\prime }(0)\rangle $ satisfies a Schr\"{o}dinger type
equation, 
\[
i\frac{\partial }{\partial s}\langle x(s)|x^{\prime }(0)\rangle =\left(
\square -\frac{R}{6}-m^{2}\right) \langle x(s)|x^{\prime }(0)\rangle \ .
\]

Assuming that $\langle x(s)|x^{\prime }(0)\rangle $ depends only on the
total geodesic distance $-(t-t^{\prime })^{2}+a^{2}\chi ^{2}$, with the
spatial part $a^{2}\chi ^{2}$ derived from Eq. (\ref{ell}), the above
equation can be solved, and we get 
\begin{equation}
\langle x(s)|x^{\prime }(0)\rangle =\frac{-i\chi }{\sinh \chi }\frac{\exp
\{im^{2}s+i[(t-t^{\prime })^{2}-a^{2}\chi ^{2}]/4s\}}{(4\pi \,s)^{2}}.
\label{nschroedinger}
\end{equation}
Substituting this solution for the integrand in Eq. (\ref{int}), gives for
the Feynman propagator 
\begin{equation}
G(x,x^{\prime })=-\frac{m}{8\pi }\frac{\chi }{\sinh \chi }\frac{%
H_{1}^{(2)}\left( m\sqrt{\left( t-t^{\prime }\right) ^{2}-a^{2}\chi ^{2}}%
\right) }{\sqrt{\left( t-t^{\prime }\right) ^{2}-a^{2}\chi ^{2}}}\ ,
\label{ff}
\end{equation}
where $H_{1}^{(2)}$ is the Hankel function of the second kind and order one.

The Hadamard function can be obtained from Eqs. (\ref{ff}) and (\ref{dfg}), 
\begin{equation}
G^{(1)}(x,x^{\prime })=\frac{m}{2\pi ^{2}}\frac{\chi }{\sinh \chi }\frac{%
K_{1}\left( m\sqrt{-\left( t-t^{\prime }\right) ^{2}+a^{2}\chi ^{2}}\right) 
}{\sqrt{-\left( t-t^{\prime }\right) ^{2}+a^{2}\chi ^{2}}},  \label{Hdrmf}
\end{equation}
where $K_{1}$ is the modified Bessel function of the second kind and order
one. The massless limit $m=0$ can be immediately be checked: 
\[
G^{(1)}(x,x^{\prime })_{m=0}=\frac{\chi }{2\pi ^{2}\sinh \,(\chi )}\left\{ 
\frac{1}{-(t-t^{\prime })^{2}+a^{2}\chi ^{2}}\right\} .
\]
Remembering that for $a\rightarrow \infty $ 
\[
\sinh \chi =a^{-1}\sqrt{(x-x^{\prime })^{2}+(y-y^{\prime })^{2}+(z-z^{\prime
})^{2}}\rightarrow \chi \ ,
\]
the well known Minkowski result is recovered for the massive and massless
cases, 
\begin{eqnarray*}
&&G^{(1)}(x,x^{\prime })=\frac{m}{2\pi ^{2}}\frac{K_{1}\left( m\sqrt{-\left(
t-t^{\prime }\right) ^{2}+r^{2}}\right) }{\sqrt{-\left( t-t^{\prime }\right)
^{2}+r^{2}}}\ , \\
&&G^{(1)}(x,x^{\prime })_{m=0}=\frac{1}{2\pi ^{2}}\left\{ \frac{1}{%
-(t-t^{\prime })^{2}+r^{2}}\right\} ,
\end{eqnarray*}
where $r$ is the geodesic distance in the spatial euclidean section.

Substituting Eq. (\ref{Hdrmf}) and the covariant derivatives (\ref{cn1}) and
(\ref{cn2}) into Eq. (\ref{Tmn}), we obtain $T(x,x^{\prime })_{\mu \nu }$.

The Klein-Gordon equation remains unchanged under isometries, 
\[
\mbox{$\pounds _\xi$}\left[ \left( \square -\frac{R}{6}-m^{2}\right) \phi %
\right] =\left( \square -\frac{R}{6}-m^{2}\right) \mbox{$\pounds _\xi$}\phi
\ , 
\]
where $\mbox{$\pounds_\xi$}$ is the Lie derivative with respect to the
Killing vector $\xi $ that generates the isometry, hence summations in the
Green functions over the discrete elements of the group $\Gamma $ is well
defined.

In ${\cal M}=H^{3}/\Gamma $, a summation over the infinite number of
geodesics connecting $x$ and $x^{\prime }$ is obtained by the action of the
elements $\gamma \in \Gamma $, which are the generators $g_{i}$ and their
products (except the identity; see below), on $x^{\prime }$. Since $\Gamma $
is isomorphic to $\pi _{1}({\cal M})$ - see section I - each geodesic
linking $x$ and $x^{\prime }$ in ${\cal M}$ is lifted to a unique geodesic
linking $x$ and $\gamma x^{\prime }$\ in $H^{3}$. Thus from Eq. (\ref{Tnl})
we get 
\begin{eqnarray}
\rho _{C} &=&\left\langle 0\left| T(x)_{\mu \nu }\right| 0\right\rangle _{%
{\cal M}}u^{\mu }u^{\nu }  \nonumber \\
&=&u^{\mu }u^{\nu }\lim_{x^{\prime }\rightarrow x}\sum_{\gamma \neq
1}T\left( x,\gamma x^{\prime }\right) _{\mu \nu }\ .  \label{resl}
\end{eqnarray}
The infinite summation occurs because the spacetime $R\times {\cal M}$ is
static, so there has been enough time for the quantum interaction of the
scalar field with the geometry to travel any distance. Since we know the
universe is expanding, the infinite summation is not physically valid. The
presence of the mass term, however, naturally introduces a cutoff.

In Eq. (\ref{resl}) the subscript $\gamma \neq 1$ means that the direct path
is not to be taken into account. We shall show, following \cite{GMM}, that
this exclusion is indeed equivalent to a renormalization of the cosmological
constant.\ \ \ \ \ 

First remember that the effective action $W$ is given by 
\[
e^{iW}=\int {\cal D}\phi \exp (iS)\ ,
\]
where the action for the scalar field can be transformed into a gaussian
type after integration by parts, 
\[
S=-\frac{1}{2}\int d^{4}x\sqrt{-g(x)}\int d^{4}y\sqrt{-g(y)}\phi
(x)F(x,y)\phi (y)\ .
\]
In a very informal way, the functional integration can be regarded as a
usual gaussian integral\footnote{%
This result is made more rigourous by assuming a particular complete
representation basis for the operator $\hat{F}$.} which, making use of Eqs. (%
\ref{okf}) and (\ref{iokf}), results in 
\begin{eqnarray*}
&&e^{iW}=\left[ \det \frac{i\hat{F}}{2\pi }\right] ^{-1/2} \\
&&W=\frac{i}{2}\ln [\det \hat{F}]+const. \\
&&\delta W=\frac{i}{2}\mbox{tr}\delta (\ln \hat{F})=\frac{i}{2}\mbox{tr}[%
\hat{F}^{-1}\delta \hat{F}] \\
&&\delta W=\frac{i}{2}\mbox{tr}\left[ i\int_{0}^{\infty }e^{-is\hat{F}%
}\delta \hat{F}ds\right]  \\
&&\delta W=\delta \left[ -\frac{i}{2}\mbox{tr}\left( \int_{0}^{\infty }\frac{%
e^{-is\hat{F}}}{s}ds\right) \right]  \\
&&W=-\frac{i}{2}\mbox{tr}\left( \int_{0}^{\infty }\frac{e^{-is\hat{F}}}{s}%
ds\right) =\int d^{4}x{\cal L}_{\mbox{{\tiny eff}}}\ .
\end{eqnarray*}
The trace of the operator is over the Hilbert space, defined in (\ref{iokf}%
), $\mbox{tr}\hat{A}=\lim_{x^{\prime }\rightarrow x}\int \sqrt{-g}%
d^{4}x\langle x|\hat{A}|x^{\prime }\rangle $. Using the Schr\"{o}dinger
kernel $\langle x|\exp (-is\hat{F})|x^{\prime }\rangle $, given in Eq. (\ref
{nschroedinger}), we obtain the effective lagrangian %\begin{equation}
%{\cal L}_{\mbox{{\tiny eff}}}=\frac{-i}{2}\sqrt{-g}\lim_{x\rightarrow
%x^{\prime }}\left[ \int_{0}^{\infty }\frac{\langle x|\exp (-is\hat{F}%
%)|x^{\prime }\rangle }{s}ds\right] =-\frac{1}{2}\sqrt{-g}\int_{0}^{\infty }%
%\frac{e^{im^{2}s}}{(4\pi )^{2}s^{3}}ds=\Lambda _{\infty }\sqrt{-g},
%\end{equation}
\begin{eqnarray}
&&{\cal L}_{\mbox{{\tiny eff}}}=\frac{-i}{2}\sqrt{-g}\lim_{x^{\prime
}\rightarrow x}\int_{0}^{\infty }\frac{\langle x|\exp (-is\hat{F})|x^{\prime
}\rangle }{s}ds  \nonumber \\
&=&-\frac{1}{2}\sqrt{-g}\int_{0}^{\infty }\frac{e^{im^{2}s}}{(4\pi )^{2}s^{3}%
}ds=\Lambda _{\infty }\sqrt{-g}\ ,
\end{eqnarray}
which shows that the direct path ($\gamma =1$), corresponds to a divergent
cosmological term $\Lambda _{\infty }$.

\section{Casimir Density {\protect\large $\protect\rho _{C}$} in a Few
Universes}

According to quantum cosmology, a smaller universe has a greater probability
of being spontaneously created. Also, the chaotic mixing becomes more
significant for smaller volumes \cite{css3}. We describe some spatially
compact universes, with increasing volumes, in subsections \ref{A}-\ref{D}.
As seen in section I, manifolds ${\cal M\cong }$ $H^{3}/\Gamma ,$ where $%
\Gamma $ is a discrete subgroup of isometries and $H^{3}$ is its universal
covering, are multiply connected.

The values of $\rho _{C}$ shown for each manifold were taken at points $%
(\theta ,\varphi )$ on the surface of a sphere inside its fundamental
region. For all of them the radius of the sphere is the same, $d=a\chi
=0.390035...a$, where $d$ is the geodesic distance. Our result is displayed
in FIGS. \ref{2}, \ref{3}, \ref{4} and \ref{5} for a scalar field with mass $%
m=0.4$, and a metric scale factor $a=10$. Angles $\theta $ and $\varphi $
correspond to the co-latitude and longitude, so the lines $\theta =0$ and $%
\theta =\pi $ correspond to the poles of the chosen sphere. For each
manifold we also write explicitly the radius of the inscribed sphere $%
R_{inradius}$.

The description that follows applies to all subsections \ref{A}-\ref{D}. The 
$g_{i}$ matrices that generate $\Gamma $ were obtained with the computer
program SnapPea \cite{sp}. The numerical code has been improved since our
previous paper \cite{dhr}. To yield a numerical result, the infinite
summation (\ref{resl}) has to be truncated. Recall that this summation
occurs in the covering space $H^{3}$. We halted the summation each time the
action of the generators $g_{i}$ and their products on the origin $%
(x^{1},x^{2},x^{3})=(0,0,0)$, reached a geodesic distance bigger than $%
d=a\chi =5.29834...a$. Care was taken so that no point was summed more than
once. In other words, the summation (\ref{resl}), which yelds $\rho _{C}$,
was truncated when the interior of the hyperbolic sphere of geodesic radius $%
d=a\chi =5.29834...a$, was covered with replicas of the fundamental region.
We checked that additional contributions were unimportant for the evaluation
of $\rho _{C}$.

\subsection{Weeks Manifold \label{A}}

\begin{figure}[tbph]
\centerline{\ \epsfxsize=4cm %\vspace{.8cm} 
\epsffile{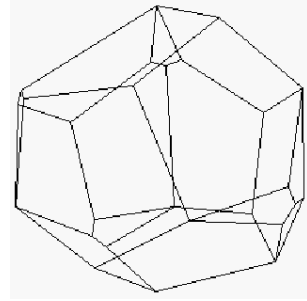}}
\caption{Fundamental region for Weeks manifold. }
\label{varw}
\end{figure}

This manifold was discovered independently by Weeks \cite{weeks} and
Matveev-Fomenko \cite{MtF}, and is the manifold with the smallest volume (in
units of $a^{3}$) known, $V=0.942707...a^{3}$. Its fundamental region is an $%
18$-face polyhedron, shown in FIG. \ref{varw}. The radius of the inscribed
sphere is $R_{inradius}=0.519162...a$.

\begin{figure}[tbp]

\centerline{\ \epsfxsize=8cm %\vspace{-1.4cm}
\epsffile{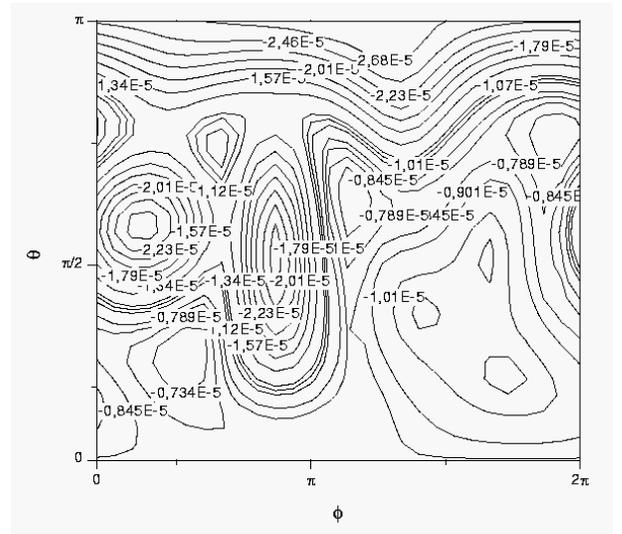}}
\caption{$\protect\rho_C$ for Weeks universe. }
\label{2}
\end{figure}

The vacuum expectation value of the $00$-component of the energy-momentum
tensor, $\rho _{C}=T_{\mu \nu }u^{\mu }u^{\nu }$, as seen by a comoving
observer, is shown in FIG. \ref{2}.

\subsection{Thurston Manifold}

\begin{figure}[tbph]
\centerline{\ \epsfxsize=4cm %\vspace{-1cm} 
\epsffile{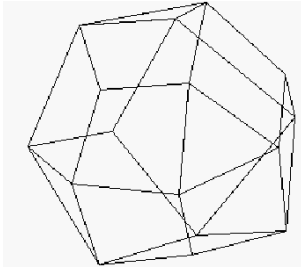}}
\caption{Fundamental region for Thurston manifold.}
\label{rfvth}
\end{figure}

It was discovered by the field medalist William Thurston \cite{Trst}. This
manifold possesses a fundamental region of $16$ faces, its volume is $%
V=0.981369...a^{3}$ (FIG. \ref{rfvth}), and $R_{inradius}=0.535437...a$.

\begin{figure}[tbph]
\centerline{\ \epsfxsize=8cm 
\epsffile{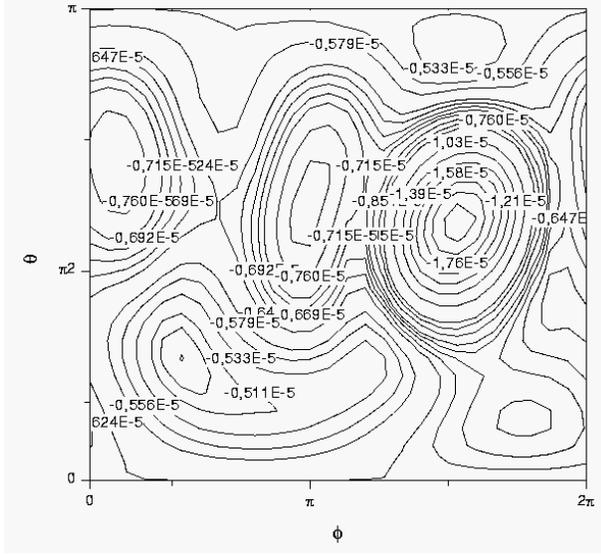}}
\caption{$\protect\rho _C$ for Thurston universe.}
\label{3}
\end{figure}

FIG. \ref{3} shows the value of $\rho _{C}=T_{\mu \nu }u^{\mu }u^{\nu
}=T_{00},$ as seen by a comoving observer. 
\subsection{Best Manifold}

\begin{figure}[tbph]
\centerline{\ \epsfxsize=4cm %\vspace{-1cm} 
\epsffile{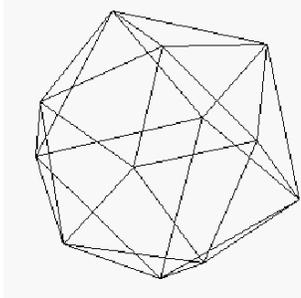}}
\caption{Fundamental region for Best manifold.}
\label{rfvb}
\end{figure}

This manifold was discovered as a by-product of a study of finite subgroups
of $SO(1,3)$ by a geometrical aproach \cite{best}. Its fundamental region is
an icosahedron with $V=4.686034...a^{3}$ and $R_{inradius}=0.868298...a$,
shown in FIG. \ref{rfvb}.

\begin{figure}[tbph]
\centerline{\ \epsfxsize=8cm %\vspace{-3.8cm} 
\epsffile{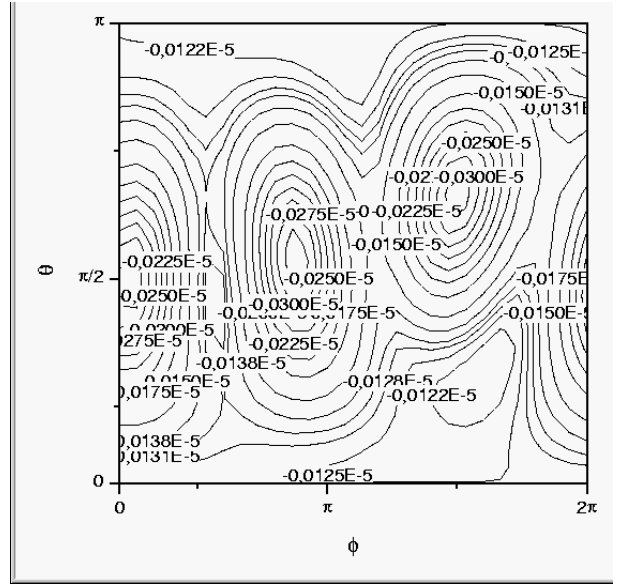}}
\caption{$\protect\rho_C$ for Best universe.}
\label{4}
\end{figure}

The vacuum expectation value of $\ \rho _{C}=T_{\mu \nu }u^{\mu }u^{\nu }$,
as seen by a comoving observer, is shown in FIG. \ref{4}.
\subsection{Seifert-Weber Manifold \label{D}}

\begin{figure}[tbph]
\centerline{\ \epsfxsize=4cm 
\epsffile{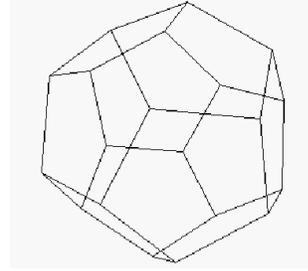}}
\caption{Fundamental region for Seifert-Weber manifold.}
\label{rfvsw}
\end{figure}

For this manifold, which was discovered by Weber and Seifert \cite{WS},  $%
V=11.199065...a^{3}$, $R_{inradius}=0.996384...a,$ and the fundamental
region is a dodecahedron (FIG. \ref{rfvsw}).

\begin{figure}[tbph]
\centerline{\ \epsfxsize=8cm  
\epsffile{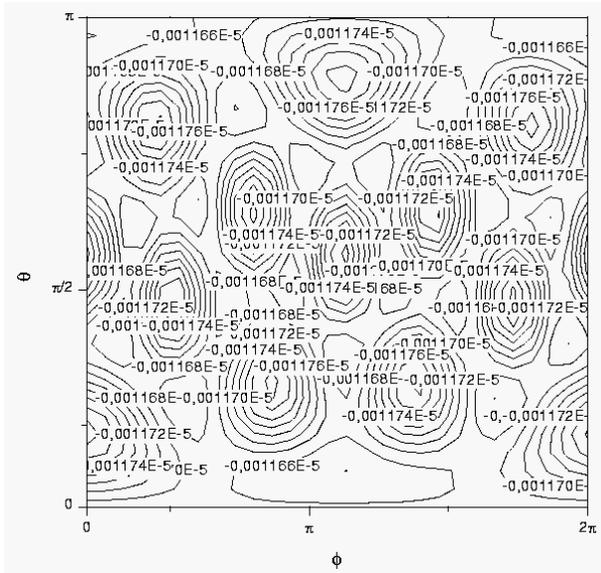}}
\caption{$\protect\rho_C$ for Seifert-Weber universe.}
\label{5}
\end{figure}

FIG. \ref{5} shows the value of $\rho _{C}=T_{\mu \nu }u^{\mu }u^{\nu },$ as
seen by a comoving observer.

\section{Conclusions}

We explicitly evaluated the distribution of the vacuum energy density of a
conformally coupled massive scalar field, for static universes with compact
spatial sections of negative curvature and increasing volume: Weeks,
Thurston, Best, and Seifert-Weber manifolds. As a specific example, we chose 
$m=0.4$ for the mass of the scalar field, and $a=10$ for the radius of
curvature. The values of the Casimir energy density $\rho _{C}$ on a sphere
of proper (geodesic) radius $d=3.90035..$ inside the fundamental polyhedron
for each of these manifolds are shown in FIGS. \ref{2}, \ref{3}, \ref{4},
and \ref{5}. In all these cases it can be seen that there is a spontaneous
generation of low multipolar components. As expected, the effect becomes
weaker for increasing volume universes.

\section*{Acknowledgments}

D.M. would like to thank the Brazilian agency FAPESP proc. no. 01/10633-3, 
for financial support.
H. V. F. thanks the Brazilian agency CNPq for partial support. R.O. thanks
FAPESP and CNPq for partial support.


\begin{references}

\bibitem{dfn2}  B. Doubrovine, S. Novikov, and A. Fomenko, {\it G\'{e}om\'{e}%
trie Contemporaine - M\'{e}thodes et Applications (}$2^{e}$ {\it Partie)},
Premi\`{e}re \'{e}dition (Mir, Moscow, 1982).

\bibitem{fagundes}  H. V. Fagundes, Astrophys. J. {\bf 291}, 450 (1985).

\bibitem{jl}  J. Levin, gr-qc/0108043 (2001).

\bibitem{LR}  M. Lachi\`{e}ze-Rey and J.-P. Luminet, Phys. Rep. {\bf 254},
135 (1995).

\bibitem{glluw}  E. Gaussman, R. Lehoucq, J. -P. Luminet, J.-P. Uzan, and J.
Weeks, Class. Quant. Grav. {\bf 18}, 5155 (2001).

\bibitem{sok}  I. Y. Sokolov, Sov. Phys. JETP Lett. {\bf 57}, 617 (1993).

\bibitem{as}  A. de Oliveira-Costa and G. F. Smoot, Astrophys. J. {\bf 448},
577 (1995).

\bibitem{rouk}  B. F. Roukema, Class. Quant. Grav. 17, 3951 (2000).

\bibitem{lll}  R. Lehoucq, M. Lachi\`{e}ze-Rey, and J.-P. Luminet, Ast. \&
Astrophys. {\bf 313}, 330 (1996).

\bibitem{llu}  R. Lehoucq, J.-P. Luminet, and J.-P. Uzan, Ast. \& Astrophys. 
{\bf 344}, 735 (1999).

\bibitem{gtrb}  G. I. Gomero, A. F. F. Teixeira, M. J. Rebou\c{c}as, and A.
Bernui, gr-qc/9811038.

\bibitem{fg}  H. V. Fagundes and E. Gausmann, Phys. Lett. A {\bf 261}, 235
(1999).

\bibitem{css2}  N. J. Cornish, D. Spergel, and G. Starkman, Class. Quant.
Grav. {\bf 15}, 2657 (1998).

\bibitem{grt}  G. I. Gomero, M. J. Rebou\c{c}as, and R. Tavakol, Class.
Quant. Grav. {\bf 18}, L145 (2001).

\bibitem{css3}  N. J. Cornish, D. Spergel, and G. Starkman, Phys. Rev. Lett. 
{\bf 77}, 215 (1996).

\bibitem{17}  N. L. Balazs and A. Voros, Phys. Rep. 143, 109 (1986).

\bibitem{lb}  J. Levin and J. D. Barrow, Class. Quant. Grav. {\bf 17}, L61
(2000).

\bibitem{dhr}  D. M\"{u}ller, H. V. Fagundes, and R. Opher, Phys. Rev. D 
{\bf 63}, 123508 (2001).

\bibitem{dh}  D. M\"{u}ller and H. V. Fagundes, gr-qc/0205050 (2002). 

\bibitem{most}  M. Bordag, U. Mohideen, V. M. Mostepanenko, Phys. Rep. {\bf %
353}, 1 (2001).

\bibitem{dfn}  B. A. Dubrovin, A. T. Fomenko, and S. P. Novikov, {\it Modern
Geometry - Methods and Applications} (Part 1), 2nd edition (Springer-Verlag,
New York, 1992).

\bibitem{Schwinger}  J. Schwinger, Phys. Rev. {\bf 82}, 664 (1951).

\bibitem{dWitt}  B. S. De Witt, Phys. Rep. {\bf 19}, 296 (1975).

\bibitem{chris1}  S. M. Christensen, Phys. Rev. D {\bf 17}, 946 (1978).

\bibitem{chris2}  S. M. Christensen, Phys. Rev. D {\bf 14}, 2490 (1976).


\bibitem{GMM}  A. A. Grib, S. G. Mamayev, and V. M. Mostepanenko, {\it %
Vacuum Quantum Effects in Strong Fields} (Friedmann Laboratory Publishing,
St. Petersburg, 1994).

\bibitem{sp}  J. Weeks, {\it SnapPea: A computer program for creating and
studying hyperbolic 3-manifolds}, freely available at site 
%TCIMACRO{\TEXTsymbol{<}}%
%BeginExpansion
\mbox{$<$}%
%EndExpansion
{\tt http://www.northnet.org/weeks}%
%TCIMACRO{\TEXTsymbol{>}}%
%BeginExpansion
\mbox{$>$}%
%EndExpansion
. \ \ 

\bibitem{weeks}  J. R. Weeks, Ph.D. thesis, Princeton University (1985).

\bibitem{MtF}  S. V. Matveev and A. T. Fomenko, Russian Math. Surveys {\bf 43%
}, No. 1, 3 (1988).

\bibitem{Trst}  W. P. Thurston, Bull. Am. Math. Soc. {\bf 6}, 357 (1982).

\bibitem{best}  L. A. Best, Can. J. Math. {\bf 23}, 451 (1971).

\bibitem{WS}  C. Weber and H. Seifert, Math. Zeitschr. {\bf 37}, 237 (1933).


\end{references}
\end{document}